# Min-Max Fine Heaps

*Suman Kumar Nath*
Department of Computer Science
University of Illinois at Urbana-Champaign
Urbana, IL 61801, USA
Email: snath@students.uiuc.edu

*Rezaul Alam Chowdhury*     *M. Kaykobad*
Department of Computer Science and Engineering
Bangladesh University of Engineering and Technology
Dhaka-1000, Bangladesh
Email: shaikat@bdonline.com

**Abstract**
In this paper we present a new data structure for double ended priority queue, called *min-max fine heap*, which combines the techniques used in fine heap and traditional min-max heap. The standard operations on this proposed structure are also presented, and their analysis indicates that the new structure outperforms the traditional one.

**Keywords:** deque, min-max heap, complexity, fine heap

## 1. Introduction

A double-ended priority queue (deque in short) is a data type supporting operations of FindMax, DeleteMax, FindMin, DeleteMin and Insertion of a new element. A traditional heap does not allow efficient implementation of all the above operations; for example, FindMin requires linear (instead of constant) time in max-heap. One approach to overcoming this intrinsic limitation of heaps is to place a max-heap "back-to-back" with a min-heap as suggested by Williams (p. 619,[6]). This leads to constant time to Find either extremum and logarithmic time to *Insert* an element or *Delete* one of the extrema, but is somewhat trickier to implement.

Min-Max heap structure, proposed by Atkinson at el.[1], overcomes these problems. The structure is based on the heap structure under the notion of min-max ordering: values stored at nodes on even (odd) levels are smaller than or equal to (respectively, greater than) values stored at their descendants. This structure can be constructed in linear time. *FindMin*, *FindMax* operations can be performed in constant time and *Insert*(*x*), *DeleteMin* and *DeleteMax* in logarithmic time using this structure. Also sub-linear merging algorithm is given with relaxation of strict ordering[3].

In this paper, we shall combine the concept of fine-heap, introduced by Carlsson[2], using bit in each node to indicate its larger child, to improve the performance of Min-Max Heap. We call it *Min-Max Fine Heap*. Also a technique similar to the one used by Gonnet and Munro[5] for traditional heaps will be employed for better performance. In the next sections we shall present the structure of the heap, algorithm for carrying out the standard operations on deque, and computational analysis of the new algorithms.

## 2. The Data Structure

Given a set *S* of values, a min-max fine heap on *S* is a binary tree *T* with the following properties:
 i. *T* has the heap-shape.
 ii. *T* is min-max ordered: values stored at nodes on even (odd) levels are smaller (greater) than or equal to the values stored at their descendants (if any) where the root is at level zero. Thus, the smallest value of S is stored at the root of *T*, whereas the largest value is stored at one of the root's children.
 iii. Each node of *T* contains a bit field that is 1 if the left child is larger than or equal to the right child, and 0 otherwise.
 An example of a min-max fine heap and its corresponding Hasse diagram are shown in figures 1 and 2.

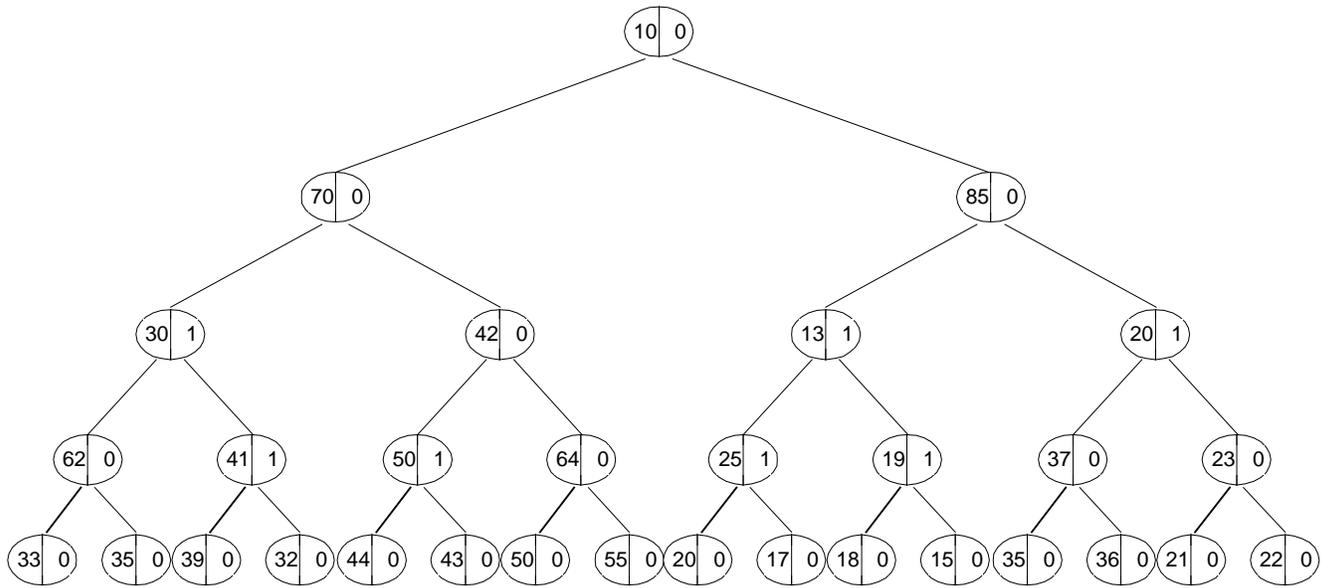

Figure 1: A Min-Max Fine Heap.

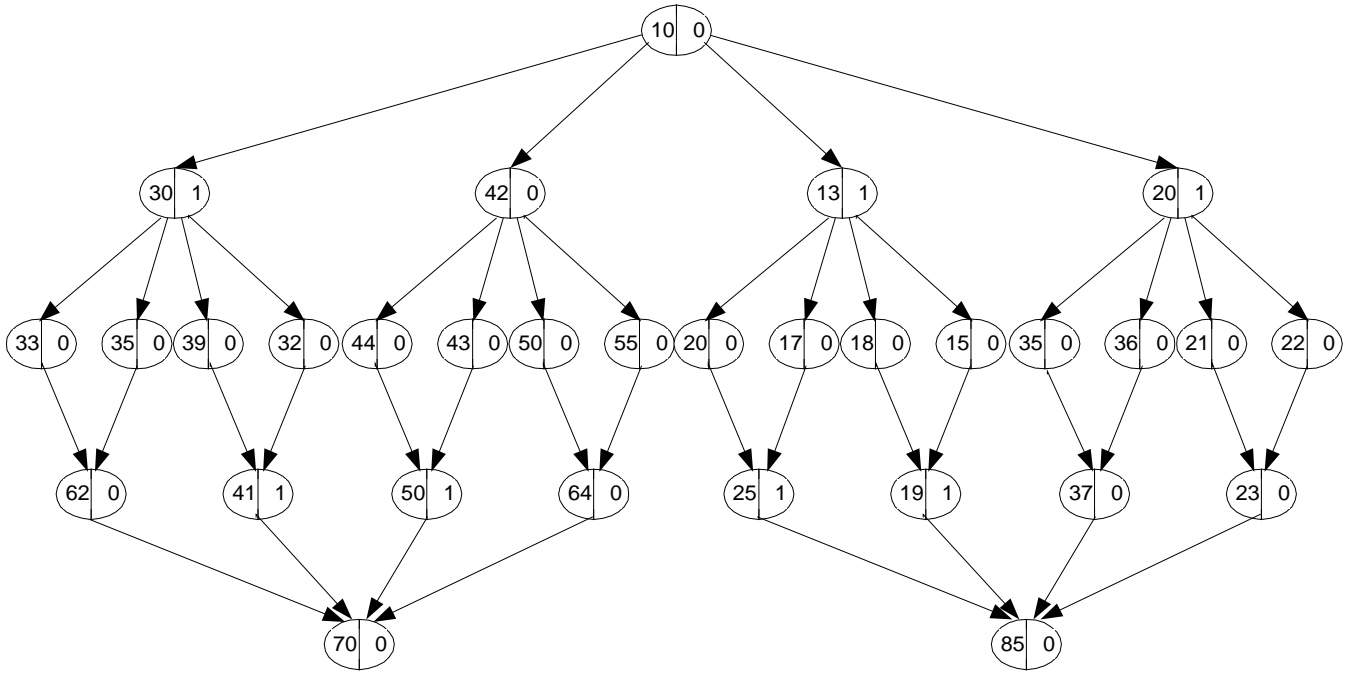

Figure 2: Hasse Diagram for Min-Max Fine Heap of Figure 1

In figure 1, the top most level is min level and the heap condition alternates between minimum and maximum from level to level. Two fields in each node of the heap and Hasse diagram indicate the value and the bit-field respectively.

## 3. Algorithms

We shall present below the algorithms for standard operations on min-max fine heap.

### 3.1 Creation

Creating a min-max fine heap is accomplished by an adaptation of Floyd's[4] linear time heap construction algorithm. It proceeds in bottom-up fashion. When the algorithm examines the subtree rooted at $A[i]$ then both subtrees of $A[i]$ are min-max ordered, whereas the subtree itself may not necessarily be min-max ordered. To adjust the element $A[i]$, we first find out a chain, using the bit information in each parent node. Referring to the Hasse diagram, the chain for inserting in max (min) level consists of the largest (smallest) immediate predecessor (ancestor) of each node, starting from the node $i$. For ease of description we define $Anc(i)$ and $Pred(i)$ to be the smallest immediate ancestor and largest immediate predecessor respectively of node $i$ in the Hasse diagram. Clearly, if $A[i]$ is on min level, the chain will be: $Anc(i)$, $Anc(Anc(i))$, $Anc(Anc(Anc(i)))$.... and for $A[i]$ on the max level, the chain will be: $Dec(i)$, $Dec(Dec(i))$, $Dec(Dec(Dec(i)))$....We then find out the appropriate position for it in the chain using binary search, insert $A[i]$ in that position by pushing up the elements in the chain above it by one step along the chain. Also, due to the push up, the bit field of some node may be destroyed. So, the bit field of the parent of the pushed-up nodes must be restored. Clearly, the algorithm must differentiate between the min- and max-levels. The resulting description of this procedure follows.

    **procedure** *TrickleDown*(*i*)
    // *i* is the position in the array
    **if** *i* is on a min level **then**
        TrickleDownMin(*i*)
    **else**
        TrickleDownMax(*i*)
    **endif**
    **end procedure**

    **procedure** *TrickleDownMin*(*i*)
    **if** $A[i]$ has children
    1. find out the nodes on the chain $C = Anc(i)$, $Anc(Anc(i))$, $Anc(Anc(Anc(i)))$.....
    2. find a position *m* in the chain such that insertion of $A[i]$ in position *m* of $C$ keeps it sorted
    3. For all the nodes *k* above the position *m* in the chain, push up the node *k* by one position along the chain and restore bit-field of parent(*k*)
    4. Place $A[i]$ in position *m* and restore the bit-field of parent(*i*)
    **end procedure**

The procedure *TrickleDownMax* is the same except that in step (1) we are to make the chain with the $Dec(i)$ nodes and in step (3), we are to push the nodes below the position *m* one position down in the chain.

### 3.2 Finding Minimum and Maximum elements

These two operations are straightforward. Root of the tree stores the minimum element and the maximum element is one of the two children of root, which is indicated by root's bit-field.

### 3.3 Insertion of an element

An element is inserted by placing it into the first available leaf position, and then reestablishing the ordering on the path from this element to the root. An efficient algorithm to insert an element can be designed by examining the Hasse diagram. The leaf-positions of the heap correspond to the nodes lying on the middle row of the Hasse diagram. To establish the min-max ordering, the new element is placed into the next available leaf position, and must then move up the diagram toward the top, or down toward the bottom, to ensure that all the paths from the top to bottom remain sorted. Thus the algorithm must first determine whether the new element should proceed further down the Hasse diagram (i.e., up the heap on max levels) or up the Hasse diagram (i.e., up the heap on successive max-levels). Once this has been determined, only grandparents along the path to the root of the heap need be examined-either those lying on min-levels or those lying on max-levels. Also, the bit fields of the parents of the displaced nodes are to be restored. But this step does not need any extra comparison, because whenever a node is moved one level above or below in the Hasse diagram, we already know its relation with the sibling of the displaced node. (If the node is moved from below (up), it must be larger (smaller) than the sibling). So updating their parent's bit field requires no extra comparison. The algorithm is as follows:

    **procedure** *BubbleUp*(*i*)
    // *i* is the position in the array
    **if** *i* is on the min level **then**
        **if** *i* has a parent **and** $A[i] > A[\text{parent}(i)]$ **then**
            make a chain $C$ with the nodes at position *i*, parent(*i*), and its successive grandparents.
        **else**
            make a chain $C$ with the nodes at position *i*, and its successive grandparents.
    **else**
        **if** *i* has a parent **and** $A[i] < A[\text{parent}(i)]$ **then**

make a chain *C* with the nodes at position *i*, parent(*i*), and its successive grandparents.
**else**
make a chain *C* with the nodes at position *i*, and its successive grandparents.
**endif**
**endif**
Find a place for *m* in the chain *C* for *A*[*i*] and inset it in position *m* by pushing the elements towards position *i* along the chain.
**end procedure**

3.4 *Deletion of minimum or maximum element*

For deleting the minimum (maximum) element, we find a chain *C* consisting of the smaller (larger) grandchildren, starting from the one of the root (larger son of root). Only if the last member of the chain also has a child, we include the child in the chain too. We do not include the last leaf during the building of the chain. We now use binary search to find an appropriate place *m* in *C* for the last leaf so that inserting it in position *m* of *C* keeps it sorted. We then move each node above position *m* in the chain to the place of its grand-parent in the heap, and then place the last leaf of the heap in position *m* in the chain, to ensure that the path running from top to bottom in the Hasse diagram remains sorted. Whenever we move a node to its grandfather's position or place the last leaf in its final position, we must update the bit-field of their new parent node. The algorithm is as follows:

**procedure** *DeleteMin*
1. MinItem ← A[root]
2. NewItem ← last leaf of the heap
3. Remove the last leaf from the heap
4. Detect the chain C consisting of smaller-grandchild(root), smaller-grandchild(smaller-grandchild(root))...
5. **if** the last member is not a leaf **then**
    include its child to the chain *C*
6. Use binary search to find the place *m* in *C*, so that inserting NewItem in position *m* in *C* keeps it sorted
7. for all nodes *k* above position *m* in the chain, in increasing order of *k*
    A[grandparent(*k*)] ← A[*k*]
    Update bit-field of parent(grandparent(k))
8. Place NewItem to the position *m* in *C*, and update its parent's bit-field.
9. return MinItem
**end procedure**

The *DeleteMax* procedure is essentially the same except that we are to start from the larger son of the root, instead of the root itself and the chain consists of smaller-grandchildren, instead of the larger ones.

## 4. Worst Case Complexity Analysis

In traditional min-max heap structure, finding the larger (smaller) child needs one comparison and finding the larger(smaller) grandchild needs three comparisons. Using bit-field in each node to indicate the larger (smaller) son reduces the number of comparison. Using bit information, we can find the larger (smaller) child of a node directly, without any comparison and the larger (smaller) grandchild of a node using only one comparison. (We can find the larger (smaller) child of a node directly from bit-field and can compare them to find their larger (smaller) child to find the larger (smaller) grandchild). This advantage improves the performance of min-max fine heap over traditional min-max heap. Some extra comparisons are needed to build up the bit fields during creation and to update bit fields of some nodes during deletion and insertion. But this extra work is overshadowed by the gain we have during finding larger (smaller) child or grandchild. In this section we shall calculate the number of comparisons required in the worst case by the proposed procedures of min-max fine heap.

*4.1 Creation*

We carry on the analysis of creation algorithm for two different types of min-max fine heaps. First, suppose the lowest level is a min-level. Suppose, at each step, we create a max level followed by a min level. So, to create a min-max fine heap with *h* max levels, we need four smaller min-max fine heaps, each with (*h*-1) max levels. So creation cost *T*(*h*) of a min-max fine heap with *h* max-levels involves the following costs:

1. Creation cost of four heaps with (*h*-1) max levels, *T*(*h*-1) for each of the heaps

*Constructing the max level*
2. Cost of finding the chain of *Pred* nodes (refer to the Hasse diagram). We need one comparison for each of the row below the middle row of the Hasse Diagram (which represents the leaf nodes), and no comparison for the levels above the middle row. So this cost becomes (*h*-1), for each of the two nodes in this max level.
3. Cost of finding a place for the nodes in the chain, using binary search. The length of the chain is 2(*h*-1) + 1, so in worst case, number of comparison is $\lceil \log(2h) \rceil$ for each of the two nodes.
4. Cost of updating the bit fields = $2(h-1)+1$, in worst case, for each of the two nodes.

*Constructing the min level*
5. Cost of finding the chain of *Anc* nodes = $h-1$
6. Cost of finding a place for the node in the chain = $\lceil \log(2h+1) \rceil$
7. Cost of updating the bit fields = $2(h-1)+2$, in worst case.

So the total cost becomes:
$T(0) = 0$
$$T(h) = 4T(h-1) + 2\lceil \log 2h \rceil + \lceil \log(2h+1) \rceil + 9h - 5$$
$$= \frac{7}{3} 4^h - 3h - \frac{7}{3} + \sum_{i=1}^{h} 4^{h-i} (\lceil \log(2h+1) \rceil + 2\lceil \log 2h \rceil)$$
$$= 1.983.n$$

Here $n = 2^{2h+1}-1$, size of the tree with $h$ max levels, where the lowest level is a min-level.

In case the lowest level is a max level, the Hasse diagram and so the analysis is somewhat different. Here, unlike the previous case, for making the first max-level we do not need any comparison. For simplicity, we calculate the cost of making two lowest levels directly, and the rest of the heap with recurrence relation.

Cost of making lowest two levels is the sum of cost of making the lowest max and min level. Making the lowest max-level involve no cost. For making each of the $2^{2h-2}$ nodes of the lowest min level, we need two comparisons. So, cost of making the lowest two levels is $2^{2h-1}$.

The recurrence relation for the rest of the tree can be deduced as before, and the relation is:

$T(1) = 0$ and,

$$T(h) = 4T(h-1) + 2\lceil \log(2h-1) \rceil + \lceil \log 2h \rceil + 9h - 10$$

After simplification, the cost becomes: $T(h) + 2^{2h-1} = 1.983...n$, same as before. Here $n = 2^{2h}-1$, size of the tree with $h$ max levels, where the lowest level is a max-level.

*4.2 Insertion*
To insert a node in a min-max fine heap with $n$ nodes, the length of the chain on which we are to perform the binary search is half the height of the tree. Thus number of comparison involved in inserting an element into a min-max fine heap with $n$ elements is $\log(\log(n+1))$. It should be noted that, during insertion, updating the bit-field of the nodes requires no extra comparison, since the relation between the two new nodes in a level is known.

*4.3 Deletion of the minimum (maximum) element*
During deletion, first we are to make a chain of smaller (larger) grandchildren. The length of the chain is half the height of the tree, and choosing each smaller (larger) grandchild requires one comparison. (To do so, we can directly determine the first child's smaller child and second child's smaller child and then one comparison between them gives the smaller grandchildren). So $0.5\log(n)$ comparisons are needed for making the chain. Binary search to find an appropriate position for the last leaf needs $\log(\log(n))$ comparisons. We also need to update bit fields of at most $0.5\log(n)$ nodes (parents of the grandchildren), and each update requires one comparison. So total number of comparison becomes $\log(n) + \log(\log(n))$

The results are compared to traditional min-heap and min-max heap in Tables 1 and 2. Here improved versions of both min-heap and min-max heap algorithms, using techniques similar to one used by Gonnet and Munro[5], are considered.

Table 1: Number of data movements

|  | Min-heap | Min-max heap | Min-max fine heap |
|---|---|---|---|
| Create | $n$ | $n$ | $n$ |
| Insert | $\log(n+1)$ | $0.5 \log(n+1)$ | $0.5 \log(n+1)$ |
| DeleteMin | $\log(n)$ | $\log(n)$ | $\log(n)$ |
| DeleteMax | $\log(n)$ | $\log(n)$ | $\log(n)$ |

Table 2: Number of comparisons

|  | Min-heap | Min-max heap | Min-max fine heap |
|---|---|---|---|
| Create | $1.625\,n$ | $2.15...n$ | $1.983...n$ |
| Insert | $\log(\log(n+1))$ | $\log(\log(n+1))$ | $\log(\log(n+1))$ |
| Delete Min | $\log(n) + g(n)$ | $1.5 \log(n) + \log(\log(n))$ | $\log(n) + \log(\log(n))$ |
| Delete Max | $0.5\,n + \log(\log(n))$ | $1.5 \log(n) + \log(\log(n))$ | $\log(n) + \log(\log(n))$ |

Tables 1 and 2 show that number of data movement is same in both min-heap and min-max fine heap, but in terms of number of comparison, min-max fine heap outperforms min-max heap.

# 5. Conclusion

Min-max fine heap we presented is basically an efficient combination of min-max heap and fine heap. The introduction of bit-field in min-max heap reduces number of comparison and thus, in spite of the overhead due to bit-fields, reduces the cost of the algorithms for it. The creation and deletion cost for the above structure is a significant improvement over min-max heap. Since the min-max pair heap structure is very much similar to a conventional min-max heap, it can also be generalized to other similar order-statistics operations more efficiently (e.g., constant time FindMedian and logarithmic time DeleteMedian).